\documentclass{ws-mpla}
\usepackage[super]{cite}
\usepackage{graphicx}
\usepackage{xcolor}
\usepackage{ulem}

\begin{document}
	
	\title{Comments on ``Numerical study of the SWKB condition of novel classes
		of exactly solvable systems''}

\author{Jonathan Bougie\footnote{jbougie@luc.edu}, Asim Gangopadhyaya\footnote{agangop@luc.edu}, Constantin Rasinariu\footnote{crasinariu@luc.edu}}
\address {Department of Physics, Loyola University Chicago, 1032 W Sheridan Rd.,\\
	 Chicago, IL 60660, U.S.A\\
 }

\maketitle
		
\begin{abstract}

We comment on the paper ``Numerical study of the SWKB condition of novel classes
of exactly solvable systems.''\cite{nasuda} We show that it  misrepresents our prior work\cite{bougie2018}, and clarify this misunderstanding.

\end{abstract}

\section{Introduction}

In ``Numerical study of the SWKB condition of novel classes of exactly solvable systems,'' \cite{nasuda} the authors (Y. Nasuda and N. Sawado) misrepresent prior literature. This includes an incorrect appendix entitled ``Appendix A. Erroneous Analysis in Ref. ~10''. We write as the authors of the aforementioned Ref. 10 in their Letter (hereafter Ref.~\refcite{bougie2018} in this Comment) to address this misrepresentation.

In Ref.~\refcite{bougie2018}, we examined a superpotential given by:
\begin{equation}
	W(x,\ell) = \frac{\omega x}2  - \frac {\ell}x + \left(\frac{2\omega x \hbar}{\omega x^2 + 2\ell -\hbar}-\frac{2 \omega x \hbar}{\omega x^2+2\ell+\hbar}\right)~, \label{eq:ExtendedSuperpotential-Quesne}
\end{equation}
equivalent to a superpotential introduced by Quesne \cite{quesne2008}.
This superpotential is an extension $W=W_0+W_h$ of the conventional radial oscillator $W_0 = \frac{\omega x}2  - \frac {\ell}x$. The conventional and extended radial oscillators share the same energy spectrum $E_n=2 n \hbar\omega$.

All conventional additive shape invariant superpotentials exactly satisfy the SWKB condition\cite{comtet1985,dutt1986,adhikari1988}:
\begin{equation}
	\int_{x_L}^{x_R} \sqrt{E_n-W^2(x)}\quad{\rm d}x = n \pi\hbar~, \quad
	\mbox{where}~ n=0,1,2,\cdots 
	\label{eq:swkb1}.
\end{equation} 
However, in Ref.~\refcite{bougie2018} we demonstrated that the SWKB condition
is not exact for the extended superpotential of Eq.~\ref{eq:ExtendedSuperpotential-Quesne}, despite its shape-invariance. 
We therefore proved ``that additive shape-invariance does not guarantee SWKB exactness by presenting a counterexample: the extended radial oscillator.'' \cite{bougie2018}. 

The authors of Ref.~\refcite{nasuda} similarly demonstrate that the SWKB condition is only approximate (not exact) for the extended radial oscillator, as well as for additional superpotentials. However, they erroneously state that the analysis in Ref.~\refcite{bougie2018} ``is wrong because it lacks the proper treatment of $\hbar$.''\cite{nasuda} In Sec.~\ref{sec:errors}, we address several misleading claims made in Ref.~\refcite{nasuda}.

\section{False and Misleading Claims of Ref. 1}\label{sec:errors}
\subsection{Scaling and use of $\hbar=1$}\label{sec:scaling}

The authors of Ref.~\refcite{nasuda} state that ``Although most of the literature employs the unit of $\hbar= 1$, to simplify the analyses, we retain $\hbar$ in this paper for rigorous discussions.''\cite{nasuda}  
However, the choice of $\hbar=1$ does not affect the rigor of the analysis  in Ref.~\refcite{bougie2018}, as we show below. 

In Ref.~\refcite{bougie2018}, we changed integration variable such that $y\equiv\sqrt{\omega} x$, so that both $y^2$ and the parameter $\ell$ have dimensions of angular momentum. With this change, the superpotential of Eq. \ref{eq:ExtendedSuperpotential-Quesne} becomes
\begin{equation}
	W(y,\ell) = \sqrt{\hbar\omega}\left[
	\frac{y}{2\sqrt{\hbar}}  - \frac {\ell}{\hbar}\frac{\sqrt{\hbar}}{y} 
	+ \frac{2 y/ \sqrt{\hbar}}{y^2/\hbar + 2\ell/\hbar - 1}
	- \frac{2 y/ \sqrt{\hbar}}{y^2/\hbar + 2\ell/\hbar + 1}
	\right]
	~. \label{eq:ExtendedSuperpotential-Quesne-y}
\end{equation}
Note that in Eq. \ref{eq:ExtendedSuperpotential-Quesne-y} the quantity in the square brackets is dimensionless. By scaling $\tilde{y}=y/\sqrt{\hbar}$ and $\tilde{\ell}=\ell/\hbar$, this becomes
\begin{equation}
	\label{eq:A}
W(\tilde{y},\tilde{\ell}) = \sqrt{\hbar\omega} \left[
	\frac{\tilde{y}}{2}  - \frac {\tilde{\ell}}{\tilde{y}}
	+ \frac{2 \tilde{y}}{\tilde{y}^2 + 2\tilde{\ell} - 1}
	- \frac{2 \tilde{y}}{\tilde{y}^2 + 2\tilde{\ell} + 1}
	\right]~.
\end{equation}
We defined $I\equiv\int_{x_1}^{x_2}\sqrt{E_n-W^2(x)}~{\rm d}x$, which can be written
\begin{equation}
	I = \hbar\int_{\tilde{y}_1}^{\tilde{y}_2} 
	\sqrt{\eta\left(\tilde{y},\tilde{\ell}\right)}\quad{\rm d}\tilde{y},
	\label{eq:integral0}
\end{equation}
where the dimensionless quantity $\eta(\tilde{y},\tilde{\ell})$ is
\begin{equation}
	\label{eq:A1}
	\eta(\tilde{y},\tilde{\ell})
	= 2n - 
		\left[
		\frac{\tilde{y}}{2}  - \frac {\tilde{\ell}}{\tilde{y}}
		+ \frac{2 \tilde{y}}{\tilde{y}^2 + 2\tilde{\ell} - 1}
		- \frac{2 \tilde{y}}{\tilde{y}^2 + 2\tilde{\ell} + 1}
		\right]^2~.
\end{equation}

Note that the quantity $I$ in Eq.(\ref{eq:integral0}) is simply $\hbar$ multiplied by a dimensionless integral; the SWKB condition we investigated was whether  $I$ equals $n\pi\hbar$. It is abundantly clear that setting $\hbar = 1$ bears no significance to the correctness of our result.

By setting $\hbar=1$ and renaming $\tilde{y}\to y$ and $\tilde\ell \to \ell$, Eq.~\ref{eq:A} above becomes identical to Eq.~26 of Ref.~\refcite{bougie2018}, and  Eq.~\ref{eq:A1} becomes equal to the expression for $\eta$ shown below Eq.~27 in Ref.~\refcite{bougie2018}. There is nothing ``devious'' about setting $\hbar=1$; it is simply a notational convenience. 

\subsection{Shape invariance and expansions in $\hbar$}

Appendix A in Ref.~\refcite{nasuda} states the following regarding Ref.~\refcite{bougie2018} (similar statements appear in the body of Ref.~\refcite{nasuda}):
\begin{quote}
	They alleged that the additive shape invariance was realized for the parameters
	$a_i$ such that $a_{i+1} = a_{i} + \hbar$. Their analysis was based on the expansion of the
	superpotential $W(a_i, \hbar)$ in power [\textit{sic}] of $\hbar$, assuming that $W$ was independent of $\hbar$
	except through the above shift of the parameter $a_i$.  The main drawback in the
	analysis was that they overlooked the dependence of the parameter $a_i$ on $\hbar$. Such
	wrong expansion with $\hbar$ inevitably leads to the devious result.\cite{nasuda}. 
\end{quote}

The statement quoted above is wrong on multiple counts. First, the shape invariance of the superpotential $W$ is not ``alleged.'' It can be verified by substituting $W$ from Eq.~(18) of Ref.~\refcite{bougie2018} (Eq.(\ref{eq:ExtendedSuperpotential-Quesne}) in these Comments) into the shape invariance condition:
			\begin{equation}
		W^2(x,a_i)  +  \hbar \frac{d\, W(x,a_i)}{dx}+g(a_i) = 
		W^2(x,a_{i+1})  -  \hbar \frac{d\, W(x,a_{i+1})}{dx}+g(a_{i+1})~,	\label{SIC1}
	\end{equation}
for parameters $a_i = \ell, \,a_{i+1}=\ell+\hbar$ (\textit{cf.} Eq.~6 of Ref.~\refcite{bougie2018}). For the superpotential of Eq.~\ref{eq:ExtendedSuperpotential-Quesne}, $g(a)=2 \omega a$.

Furthermore, contrary to the claim of Ref.~\refcite{nasuda}, the analysis of Ref.~\refcite{bougie2018} is not based on $\hbar$-expansion. The authors of Ref.~\refcite{nasuda} misunderstood the scope of  the $\hbar$-expansions discussed in  ``Sec.~1: Introduction'' and ``Sec.~2: Preliminaries,'' of Ref.~\refcite{bougie2018}. These expansions simply placed
our work in the context of the existing literature. Specifically, subsection 2.2 illustrates that Quesne's superpotential is a solution of previously derived partial differential equations \cite{bougie2010,bougie2012}. The status of this superpotential as a valid shape-invariant superpotential
can be verified by direct substitution in Eq. (\ref{SIC1}), independent of any $\hbar$ expansion.

The main result of  Ref.~\refcite{bougie2018} was to prove that the extended shape-invariant superpotential given by Quesne \cite{quesne2008} is not SWKB-exact via numerical integration. The expansion played no role in the new results obtained in  Ref.~\refcite{bougie2018}.

\subsection{Dimension of the shape-invariance parameter}

Finally, the authors of Ref.~\refcite{nasuda} claim that in Ref.~\refcite{bougie2018} ``the system the authors considered is irrelevant to the known quantum mechanical problems such as the well-known radial oscillator for the explicit factor $\hbar$.''\cite{nasuda} 

They are wrong here as well. In Ref.~\refcite{bougie2018}, we consider shape invariant parameters $a_{i+1}=a_i+\hbar.$ Therefore, the parameter $\ell$ has dimensions of angular momentum as discussed in Sec~\ref{sec:scaling} above.
The authors of Ref.~\refcite{nasuda} dedicate much of their Appendix A to show the trivial result that $\ell=\hbar(\ell'+1)$, where $\ell'$ corresponds to a dimensionless quantum number. This correspondence does not invalidate any results in Ref.~\refcite{bougie2018} and we did not overlook it (\textit{cf.} Footnote 3 in Ref.~\refcite{bougie2018}).

\section{Conclusion}

The authors of Ref.~\refcite{nasuda} appear to have misunderstood the existing scientific literature. The claim of Ref.~\refcite{bougie2018} was to ``present a concrete example of an additive shape invariant
	potential for which the SWKB method fails to produce exact results.'' \cite{bougie2018}. The superpotential used in Ref.~\refcite{bougie2018} is additive shape-invariant, and the numerical analysis indeed demonstrated its SWKB-inexactness. The analysis of Ref.~\refcite{bougie2018} is correct. 

\bibliographystyle{ws-mpla}

\end{document}